\documentclass[preprint,nofootinbib,amsmath,prd,aps,superscriptaddress,tightenlines,12pt]{revtex4}
\usepackage{graphicx}
\usepackage{graphics}
\usepackage{amsmath}
\usepackage{hyperref}
\usepackage{xcolor}
\usepackage{xspace}
\usepackage{subfigure}

\def\lsim{\mathrel{\rlap{\lower4pt\hbox{\hskip1pt$\sim$}}
    \raise1pt\hbox{$<$}}}                
\def\gsim{\mathrel{\rlap{\lower4pt\hbox{\hskip1pt$\sim$}}
    \raise1pt\hbox{$>$}}}                

\def\OMIT#1{}

\newcommand{\be}{\begin{eqnarray}}
\newcommand{\ee}{\end{eqnarray}}

\newcommand{\nn}{\nonumber}
\newcommand{\w}{\omega}

\newcommand{\bea}{\begin{eqnarray}}
\newcommand{\eea}{\end{eqnarray}}

\newcommand{\bnp}{\bar n \!\cdot\! p}

\newcommand{\pTcut}{p_T^{\rm cut}}
\newcommand{\pTjet}{p_T^{\rm jet}}

\newcommand{\pTZ}{p_T^{\rm Z}}

\def\drlj{\ensuremath{\Delta R^{\ell j}}}
\def\drll{\ensuremath{\Delta R^{\ell\ell}}}

\def\mll{\ensuremath{m^{\ell\ell}}}
\def\gev{\ifmmode {\mathrm{\ Ge\kern -0.1em V}}\else
                   \textrm{Ge\kern -0.1em V}\fi}%

\def\lsim{\mathrel{\rlap{\lower4pt\hbox{\hskip1pt$\sim$}}
    \raise1pt\hbox{$<$}}}                
\def\gsim{\mathrel{\rlap{\lower4pt\hbox{\hskip1pt$\sim$}}
    \raise1pt\hbox{$>$}}}                

\def\OMIT#1{}

\begin{document}

\setlength\baselineskip{17pt}

\begin{flushright}
\vbox{
\begin{tabular}{l}
\end{tabular}
}
\end{flushright}
\vspace{0.1cm}


\title{\bf Jet vetoes versus giant K-factors in the exclusive Z+1-jet cross section}

\vspace*{1cm}

\author{Radja Boughezal}
\email[]{rboughezal@anl.gov}
\affiliation{High Energy Physics Division, Argonne National Laboratory, Argonne, IL 60439, USA} 
\author{Christfried Focke}
\email[]{christfried.focke@gmail.com}
\affiliation{Department of Physics \& Astronomy, Northwestern University, Evanston, IL 60208, USA}
\author{Xiaohui Liu}
\email[]{xhliu@umd.edu}
\affiliation{Maryland Center for Fundamental Physics, University of Maryland, College Park, Maryland 20742, USA} 
\affiliation{Center for High-Energy Physics, Peking University, Beijing, 100871, China}


  \vspace*{0.3cm}

\begin{abstract}
  \vspace{0.5cm}
  
We perform a detailed study of the exclusive $Z$+1-jet cross section at the 13 TeV LHC, motivated by the importance of similar 
exclusive cross sections in understanding the production of the Higgs boson in the $W^+W^-$ final state.  We point out a feature of the 
ATLAS analysis that has significant impact on the theoretical predictions: the jet-isolation criterion implemented by ATLAS effectively allows dijet events where an energetic jet is collinear to a final-state lepton.  This process contains a giant $K$-factor arising from the collinear emission of a $Z$-boson from the dijet configuration.  This overwhelms the effect of the jet-veto logarithms, making it difficult to test their resummation in this process.   We provide numerical results that demonstrate the interplay between the jet-veto logarithms and the giant $K$-factor in the theoretical prediction.  We study several observables, including the transverse momentum distributions of the leading jet and the $Z$-boson, in the exclusive $Z$+1-jet process, and discuss their sensitivity to both the giant $K$-factor and the jet-veto logarithms.  We suggest a modified isolation criterion that removes the giant $K$-factor and allows for a direct test of the jet-veto resummation framework in the exclusive $Z$+1-jet process.

\end{abstract}

\maketitle

\section{Introduction}
\label{sec:intro}

The study of QCD cross sections in the presence of exclusive jet binning has received significant theoretical attention over the past few years.  This interest is driven primarily by the experimental need to separate signal from background in the search for the Higgs boson in the $W^+W^-$ final state~\cite{Aad:2012tfa,Aad:2012uub,Chatrchyan:2012ufa,Chatrchyan:2012ty}.  This analysis proceeds by separating the 0-jet and 1-jet bins from the inclusive 2-jet bin, where the $t\bar{t}$ background contamination is large. This separation allows for different cuts to be imposed in the inclusive 2-jet bin to reduce the $t\bar{t}$ background.

Predictions in fixed-order perturbation theory in the presence of exclusive jet binning can suffer from large uncertainties, due in part to unresummed logarithms involving the disparate scales in the process~\cite{Berger:2010xi,Stewart:2011cf,Gangal:2013nxa}.  The large logarithms in question take the form $L\,=\, \ln\left(  Q/p_T^{cut}\right)$, where $Q$ is the hard scale of the considered process and $\pTcut$ denotes the upper cut on the transverse momentum of additional final-state jets.  These terms can also shift the central value of the prediction.  Such an effect has recently been invoked~\cite{Meade:2014fca,Jaiswal:2014yba,Monni:2014zra} to explain the slight excess in the $WW$ cross section compared to theoretical predictions observed by ATLAS and CMS~\cite{ATLASWW,CMSWW}.

It is now known how to resum these logarithms to all orders in the QCD coupling constant, both in the 0-jet bin~\cite{Stewart:2013faa,Banfi:2012yh, Banfi:2012jm,Becher:2012qa,Banfi:2013eda,Becher:2013xia,Tackmann:2012bt,Gangal:2014qda} and the 1-jet bin~\cite{Liu:2012sz,Liu:2013hba,Boughezal:2013oha,Boughezal:2014qsa}.  A combined treatment of the 0-jet and 1-jet bins indicates that a factor of two reduction in the theoretical uncertainty on the prediction for Higgs production in the $WW$ final state is possible upon switching from fixed-order to resummation-improved perturbation theory~\cite{Boughezal:2013oha}.  This uncertainty will be further reduced given completion of the full next-to-next-to-leading order calculation of the Higgs+1-jet cross section~\cite{Boughezal:2013uia,Chen:2014gva}.  Given the potential impact of the resummation framework on reducing the theoretical uncertainty in the presence of exclusive jet binning, it is highly desirable to test it against experimental data in processes not involving the Higgs boson.  Candidate processes for this test should have a large rate, feature a clean experimental signature, and possess  a large hierarchy between the scales $Q$ and $p_T^{cut}$.  Two obvious choices that fulfill these criteria are the exclusive $W$+jet and $Z$+jet processes.  The possibility of measuring the spectrum over a wide range of the jet transverse momentum, $\pTjet$, or the $Z$-boson transverse momentum $\pTZ$, allows the logarithms $L$ in the theoretical prediction to be probed over a wide range of values, since $Q \sim \pTjet,\pTZ$.  

In this manuscript we perform a detailed study of the $Z$+1-jet process at the 13 TeV LHC.  We discuss the kinematics in detail, study the effect of higher-order corrections on several distributions, and investigate the impact of the resummation of jet-veto logarithms on the exclusive 1-jet bin.   Looking at the ATLAS analysis presented in~\cite{Aad:2013ysa} for this process, we have identified a critical aspect of their isolation requirement that affects the selection of the $Z$+1-jet events.  The ATLAS analysis effectively accepts two-jet events where a $Z$-boson is collinear to a final-state jet. In the first step of the experimental analysis such events are vetoed by a combination of isolation requirements and lepton-quality cuts.  However, the measured cross section is then extrapolated using Monte-Carlo simulation to include this collinear phase-space region in the unfolding scheme implemented by ATLAS.  The ATLAS measurements using the current isolation requirement should therefore be thought of as the sum of two distinct processes: a $Z$+1-jet exclusive cross section with a global jet veto imposed, and a dijet cross section with the emission of a $Z$-boson within a small cone surrounding one of the two jets.   At high $\pTjet$ the second process leads to a ``giant K-factor"~\cite{Rubin:2010xp}.
The second process becomes large at high $\pTjet$ due to the turn-on of new, large partonic scattering processes.
We study the effect of such a large $K$-factor on the fixed-order cross section at high $\pTjet$ values for the accessible phase-space region at 13 TeV and its interplay with the  jet-veto resummation in the theoretical prediction. Since the analogous Higgs+1-jet process does not receive contributions from similar new, large scattering processes at higher orders, it is desirable to reduce their effect while maintaining sensitivity to the jet-veto resummation.  We therefore suggest an alternative isolation criterion that removes the giant $K$-factor effect and selects only the $Z$+1-jet events.  We also study the effect of jet-veto resummation on the $\pTZ$ distribution, for which the sensitivity to giant $K$-factors is reduced.

Our paper is organized as follows. In Section~\ref{sec:atlas} we describe the cuts imposed in the ATLAS measurement, paying careful attention to the isolation requirement on the leptons.  In section~\ref{sec:kinematics} we study in detail various kinematic observables
using the ATLAS isolation criteria. In Section~\ref{sec:theory} we discuss the framework we use for our theoretical predictions.  Numerical results for 13 TeV LHC collisions are presented in Section~\ref{sec:numerics}. 
We discuss the structure of the numerical results in detail, show how the alternate isolation proposed in Section~\ref{sec:atlas} allows the jet-veto resummation framework to be directly tested, and consider also the use of the $Z$-boson transverse momentum distribution.  Finally, we conclude in Section~\ref{sec:conc}.
 
 \section{Overview of the ATLAS measurement} 
\label{sec:atlas}

We begin with a discussion of the experimental cuts used in the 7 TeV ATLAS measurement of the exclusive $Z$+1-jet cross section~\cite{Aad:2013ysa}, which we will use as a template for a 13 TeV analysis.  We also propose an alternative isolation requirement that should be possible to implement experimentally.  As we will discuss later, our suggestion reduces the effect 
of large new partonic scattering channels that occur first at next-to-leading order (NLO) in QCD perturbation theory.  This consequently increases the sensitivity of this process to the resummation framework for jet-veto logarithms, which we wish to test using this process.
 
The ATLAS selection criteria on both the jets and leptons are summarized in Table~\ref{t:selection}.  We note that we use the combined data sample that includes the $Z$-boson decays to electrons and muons as provided in~\cite{Link2ATLASdata}.  This sample is constructed by extrapolating the slightly different cuts on the electrons and muons to a common phase-space region.  Jets are reconstructed using the anti-$k_t$ algorithm~\cite{Cacciari:2008gp} with a distance parameter $R=0.4$.  The jet candidates are required to have a transverse momentum $\pTjet>30\>{\rm GeV}$, a rapidity cut $|y^{{\rm jet}}|<4.4$ and a lepton-jet separation $\Delta R^{lj}>0.5$.  The leptons are required to satisfy $p_{T}^l>20\>{\rm GeV}$ and to have a pseudo-rapidity in the range $|\eta^l|<2.5$.
 
 \begin{table}[h!]
\centering
\begin{tabular}{|l|l|l|}
\hline\hline
lepton $p_T$\         &  \multicolumn{2}{|c|}{$p_T^l >20\gev$} \\
lepton $|\eta |$  & \multicolumn{2}{|c|}{$|\eta^l|<2.5$} \\
 \hline
lepton  charges     & \multicolumn{2}{|c|}{opposite charge}\\
lepton separation  \drll\    & \multicolumn{2}{|c|}{$\drll >0.2$ }\\
lepton invariant mass \mll       & \multicolumn{2}{|c|}{$66\gev \leq \mll \leq 116\gev$ }\\ 
\hline\hline
jet $p_T$\               & \multicolumn{2}{|c|}{$\pTjet >  30\gev$ }\\
jet rapidity  $y^{{\rm jet}}$\     & \multicolumn{2}{|c|}{$ |y^{{\rm jet}}|<4.4$ }\\
lepton-jet separation \drlj\    & \multicolumn{2}{|c|}{$\drlj >0.5$ }\\
\hline\hline
\end{tabular}
\caption{Summary of the $Z \to ll $\  and jet selection criteria.~\label{t:selection}}
 
\end{table}
 
An important issue in this analysis is the implementation of lepton-jet isolation.  In the ATLAS analysis, the two leptons and jet are required to satisfy the isolation criteria $\Delta R^{lj}\geq 0.5$.  However, the experimental measurement is inclusive in hadronic activity inside the cones around each lepton.  Although jets collinear to leptons are initially removed by lepton quality and isolation constraints, a Monte-Carlo unfolding correction is later implemented in the analysis that removes the effect of these cuts.  This has the consequence that no event veto is imposed on an energetic jet that falls within either cone.\footnote{We thank Joey Huston for discussions on this point.}  Events with two energetic jets, with one jet collinear to a lepton, are therefore accepted by the ATLAS analysis.  We note also that to match the theoretical predictions from Blackhat+Sherpa~\cite{blackhat} to which ATLAS compares their 7 TeV results, such events must be included.\footnote{We thank Daniel Maitre for confirming the theoretical predictions from Blackhat+Sherpa.}  We will combine the effect of experimental cuts and the ATLAS unfolding scheme into an effective ATLAS isolation requirement in which collinear jets are kept.  As we will show later, these events with a jet collinear to a lepton lead to the appearance of a giant $K$-factor at high $\pTjet$ due to the emission of a $Z$-boson collinear to a very energetic final-state jet.  We define and later study an alternative lepton-jet isolation criterion that instead vetoes energetic jets that satisfy $\Delta R^{lj}\leq 0.5$.  This could be implemented in the experimental analysis by modifying the unfolding scheme.  The alternate isolation requirement has the effect of removing the giant $K$-factor.  To summarize, we consider the following two isolation criteria in our study:\\\\
$-${\bf \underline{ATLAS isolation}}: jets with $\Delta R^{lj}\leq 0.5$ are kept;
\\
$-${\bf \underline{Alternative isolation}}: jets with $\Delta R^{lj}\leq 0.5$ are vetoed.\\

Finally, we note that ATLAS applies corrections to the theoretical predictions in their comparison to fixed-order QCD to account for the underlying event and for QED final-state radiation effects.  The effect of the underlying event is found to be roughly 7\% at low $\pTjet$, falling to zero at high $\pTjet$~\cite{Aad:2013ysa}.  The QED final-state radiation correction factor is an additional 2\%. Since we are interested primarily in the high-$\pTjet$ region in our analysis, and since most of our study involves comparing the fixed-order results to the resummed ones which receive the same shifts, we neglect these corrections here.  
 
\section{Kinematical considerations} 
\label{sec:kinematics}
We have pointed out that the ATLAS isolation criterion leads to the acceptance of events with a second jet collinear to a lepton, and the appearance of a giant $K$-factor at high $\pTjet$ due to these dijet events.  The dominant kinematic configuration that contributes to these corrections comes from a boosted $Z$-boson collinear to a final-state jet. The leptons produced are collimated along the $Z$-boson direction. Since this underlying kinematical picture motivates the alternate isolation criterion described above, we provide here various numerical investigations that support this assertion.  This will also provide us with a picture of what a typical scattering event looks like at high jet transverse momentum.  We focus on the leading-$\pTjet$ region of $1600$ GeV$<\pTjet<2000$ GeV for $\sqrt{s}=13$ TeV, the highest bin in which we expect an appreciable number of events after the running of a high-luminosity LHC. 
We study three observables to gain intuition into a typical scattering event.
\begin{itemize}
\item We study the differential cross section as a function of the separation between the $Z$-boson and the jets using the standard distance measure 
$\Delta R=\sqrt{(\Delta\eta^2+\Delta\phi^2)}$. 

\item We also study the differential cross section as a function of the $Z$-boson $p_T$. 

\item Finally, we study the separation between the $Z$-boson and the leptons. We will show the result as a function of both the maximum and minimum separation between
the $Z$-boson and the leptons. 
\end{itemize}

We begin by discussing the $\Delta R_{Zj}$ distribution.  If the $Z$-boson is predominantly collinear, we should see a peak near $\Delta R_{Zj}=0$. If it is predominantly soft, we should see a peak near small values of the $Z$-boson $p_T$. In Table~\ref{t:Zjcoll}, we show the results of the scans in the form of bin-integrated cross sections.  All cross sections are at NLO in fixed-order perturbation theory, using the ATLAS isolation criterion.
\begin{table}[h!]
\centering
\begin{tabular}{|l|l|l|l|}
\hline\hline
&\multicolumn{2}{|c|}{$\sigma^{NLO}\ [fb]$} & Ratio \\
\hline
$0.0< \Delta R_{Zj} <0.5$ & \multicolumn{2}{|c|}{$0.0231$} & 0.796 \\
$0.5< \Delta R_{Zj} <1.0$ & \multicolumn{2}{|c|}{$0.0051$} & 0.174 \\
$1.0< \Delta R_{Zj} <1.5$ & \multicolumn{2}{|c|}{$0.0006$} & 0.020 \\
\hline\hline
\end{tabular}
\caption{The differential cross section as a function of the separation between the $Z$-boson and the jets, $\Delta R_{Zj}$, for 
$p_T^Z>40$~GeV, $1600$~GeV$<p_T^{jet}<2000$~GeV and $\sqrt{s}=13$ TeV. The last column shows the ratio of the
result in each bin to the total NLO cross section~\label{t:Zjcoll}
}
\end{table}
The cross section is clearly larger for the collinear region $0.0< \Delta R_{Zj} <0.5$; approximately 80\% of the events fall into this bin, indicating that most events do have a $Z$-boson collinear to a jet.  

We now impose a lower cut 
on the $p_T$ of the $Z$-boson, and scan over the value of the lower cut. We again focus on the $\pTjet$ range $1600$~GeV$<\pTjet<2000$~GeV, with no cut on $\Delta R_{Zj}$.
\begin{table}[h!]
\centering
\begin{tabular}{|l|l|l|}
\hline\hline
$p_{T,min}^Z$ & \multicolumn{2}{|c|}{$d\sigma^{NLO}\ [fb]$} \\
\hline
$40$ & \multicolumn{2}{|c|}{$0.02923$} \\
$80$ & \multicolumn{2}{|c|}{$0.02869$} \\
$100$ & \multicolumn{2}{|c|}{$0.02833$} \\
$200$ & \multicolumn{2}{|c|}{$0.02555$} \\
$400$ & \multicolumn{2}{|c|}{$0.01766$} \\
$500$ & \multicolumn{2}{|c|}{$0.01413$} \\
$1000$ & \multicolumn{2}{|c|}{$0.00263$} \\
$1500$ & \multicolumn{2}{|c|}{$0.00031$} \\
\hline\hline
\end{tabular}
\caption{A scan of the differential cross section as a function of the $p_{T}^Z$, for $1600$~GeV$<p_T^{jet}<2000$~GeV and $\sqrt{s}=13$ TeV.~\label{t:Zsoft}}
\end{table}
The results in table~\ref{t:Zsoft} are nearly identical for the regions $p_T^Z>40$~GeV through $p_T^Z>200$~GeV, and only decrease significantly when 
$p_T^Z>500$~GeV.  This indicates that over 85\% of the events have $p_T^Z>200$~GeV, and that 60\% of the events have $p_T^Z>400$~GeV.  
While the $Z$-boson is therefore not too soft for most of the events, in general $\pTjet > \pTZ$.  This will have implications later when we discuss how $\pTjet$ and $\pTZ$ behave for the two isolation criteria. 

Finally, we study the separation between the $Z$-boson and the leptons, looking at events in each $ \Delta R_{Zl}$ bin as a function
of both the maximum and the minimum distances between the $Z$-boson and the two leptons, for $1600$~GeV$<\pTjet<2000$~GeV.
 \begin{table}[h!]
\centering
\begin{tabular}{|l|l|l|l|}
\hline\hline
&\multicolumn{2}{|c|}{$d\sigma^{NLO}\ [fb]$} & Ratio \\
\hline
$0.0< \Delta R_{Zl}^{max} <0.5$ & \multicolumn{2}{|c|}{$0.02006$} & 0.706 \\
$0.5< \Delta R_{Zl}^{max}  <1.0$ & \multicolumn{2}{|c|}{$0.00708$} & 0.249 \\
$1.0< \Delta R_{Zl}^{max} <1.5$ & \multicolumn{2}{|c|}{$0.00708$} & 0.044 \\
\hline\hline
\end{tabular}
\caption{A scan of the differential cross section as a function of the maximum distance between the $Z$ and the leptons $\Delta R_{Zl}^{max}$ for  $1600$~GeV$<p_T^{jet}<2000$~GeV and for 
$\sqrt{s}=13~TeV$. The last column shows the ratio of the result in each bin to the total NLO cross section~\label{t:Zlcollmax}}
\end{table}
\begin{table}[h!]
\centering
\begin{tabular}{|l|l|l|l|}
\hline\hline
&\multicolumn{2}{|c|}{$d\sigma^{NLO}\ [fb]$} & Ratio \\
\hline
$0.0< \Delta R_{Zl}^{min}  <0.5$ & \multicolumn{2}{|c|}{$0.0284$} & 1.000 \\
$0.5< \Delta R_{Zl}^{min}  <1.0$ & \multicolumn{2}{|c|}{$0.0$} & 0.000 \\
\hline\hline
\end{tabular}
\caption{A scan of the differential cross section as a function of the minimum distance between the $Z$ and the leptons $\Delta R_{Zl}^{min}$ for  $1600$~GeV$<p_T^{jet}<2000$~GeV and for 
$\sqrt{s}=13~TeV$. The last column shows the ratio of the result in each bin to the total NLO cross section~\label{t:Zlcollmin}}
\end{table}
As is shown in both tables~\ref{t:Zlcollmax} and~\ref{t:Zlcollmin}, most events have $0.0< \Delta R_{Zl}<0.5$, indicating that the leptons are collimated with the $Z$-boson direction, as claimed.
 
These results together indicate that the typical event predicted by NLO perturbation theory for the ATLAS experimental setup possesses a relatively high transverse momentum $Z$-boson emitted close to the highest-$p_T$ jet.

\section{Theoretical framework}
\label{sec:theory}

We now discuss the theoretical framework we use to provide our predictions.  The initial expectation for the high-$\pTjet$ exclusive $Z$+1-jet bin is that it consists of a high-$p_T$ $Z$-boson back-to-back in the transverse plane from the jet.  The kinematical considerations in the previous section make it clear that this expectation is too naive.  The ATLAS isolation criterion is such that the accepted cross section consists of two distinct categories of events: exclusive $Z$+1-jet events with a global jet veto imposed, and dijet events where one jet is collinear to a final-state lepton.  This motivates the following theoretical decomposition of the cross section:
\begin{equation}
\sigma_{\text{total}} = \sigma_{\text{Z+1j}}+\sigma_{\text{dijet}}.
\end{equation}
Our theoretical formalism allows us to resum the jet-veto logarithms that appear in $\sigma_{\text{Z+1j}}$. The contribution from $\sigma_{\text{dijet}}$ is obtained by matching our resummation prediction to the fixed-order result at next-to-leading order (NLO) in $\alpha_s$, the first order at which $\sigma_{\text{dijet}}$ appears.

We begin with a brief sketch of the resummation framework.  This formalism has already been discussed extensively in the literature.  For a more detailed treatment of the exclusive 1-jet bin resummation we refer the reader to Refs.~\cite{Liu:2012sz,Liu:2013hba}.  The measurement function for $\sigma_{\text{Z+1j}}$ consists of a single jet with $\pTjet > \pTcut$, together with a $Z$-boson which decays leptonically.  A global veto over all of phase space is imposed on any other jet with $\pTjet > \pTcut$.  Since $\pTcut$ is substantially lower than the partonic center-of-mass energy, the cross section is sensitive to soft and collinear emissions, leading to large logarithms in the prediction.  The resummation of jet-veto logarithms begins with the factorization of the cross section into separate hard, soft, and collinear sectors which follows from this hierarchy of scales.  We use soft-collinear effective theory (SCET) to accomplish this factorization~\cite{Bauer:2000ew,Bauer:2000yr,Bauer:2001ct,Bauer:2001yt,Bauer:2002nz}.  The detailed steps in the derivation are presented in Ref.~\cite{Liu:2012sz}.  We present here only the final result for the factorized cross section:
\bea\label{factgen}
\mathrm{d}\sigma^{\text{NLL}^{\prime}}_{\text{Z+1j}} &=& \mathrm{d}\Phi_{Z}\mathrm{d}\Phi_{J}\,
{\cal F}(\Phi_{Z},\Phi_{J})
\,
\sum_{a,b}\int \mathrm{d}x_{a} \mathrm{d}x_b \frac{1}{2\hat{s}}\,
 (2\pi)^4 \delta^4\left(q_a + q_b - q_{J} -q_{Z}\right)\nn\\
&&\times 
\bar{\sum_{\rm spin}}
\bar{\sum_{\rm color}}
{\rm Tr}(H\cdot S)\,
{\cal I}_{a,i_aj_a} \otimes f_{j_a}(x_a)\,
{\cal I}_{b,i_bj_b} \otimes f_{j_b}(x_b)
J_{J}(R)\,.
\eea
The superscript on the differential cross section indicates that we will evaluate this cross section to the $\text{NLL}^{\prime}$ level, in the counting scheme defined in Ref.~\cite{Berger:2010xi}.  $\mathrm{d}\Phi_{Z}$ and $\mathrm{d}\Phi_{J}$
are the phase-space measures for the $Z$-boson  and 
the massless jet $J$, respectively. ${\cal F}(\Phi_{Z},\Phi_{J})$ includes all additional
phase-space cuts other than the transverse momentum veto.  $H$ is the hard function that 
comes from matching QCD onto SCET.  In the scheme in which we work, the hard function is the finite part of the one-loop virtual corrections to the $Z$+1-jet amplitude.  $S$ describes soft final-state emissions.  The trace is over the color indices.  The functions ${\cal I}$ and $J$ describe collinear emissions along the beam axes and along the final-state jet direction, respectively.  The measured transverse momentum of the leading jet $\pTjet$ should be much larger than $\pTcut$.  Implicit in the above setup is that the dominant kinematic configuration leading to the final state is a hard $Z$-boson recoiling against a hard jet.

The functions $H$, $J$, $B$ and $S$ all live at different energies, in the sense that the large logarithms they contain are minimized by different scale choices.  However, each function obeys a separate renormalization group equation that allows it to be evolved to its natural scale, thereby resumming the large logarithms.  The requisite anomalous dimensions, as well as the one-loop jet, beam and soft functions needed for a full $\text{NLL}^{\prime}$ result, are given in Refs.~\cite{Liu:2012sz,Liu:2013hba}.  They are reproduced for completeness in the Appendix of this manuscript.  The one-loop hard functions can be obtained from Ref.~\cite{Becher:2012xr,Gehrmann:2011ab}.  

The final ingredient needed for our result is the matching of the resummed cross section with the fixed-order NLO result. We use the NLO predictions for $Z$+1-jet contained in MCFM~\cite{Campbell:2010ff}.  We obtain our prediction by setting
\bea\label{rgimproved}
\sigma^{\text{NLL}^{\prime}+\text{NLO}}_{\text{Z+1j}}= \sigma^{{\rm NLL}'}_{\text{Z+1j}} + \,
\sigma^{\rm NLO}_{\text{Z+1j}}-\sigma^{{\rm NLL',exp}}_{\text{Z+1j}}.
\eea
In this equation, $\sigma^{\rm NLO}$ is the fixed-order NLO 
cross section obtained from MCFM, and $\sigma^{\rm NLL'}$ is the resummed 
cross section up to $\text{NLL}^{\prime}$ accuracy presented in Eq.~(\ref{factgen}).  $\sigma^{{\rm NLL',exp}}_{\text{Z+1j}}$
captures the singular features of $\sigma^{\rm NLO}$, and is obtained by expanding 
$\sigma^{\rm NLL'}$ in $\alpha_s$ with all scales set to a common value $\mu=H_T/2$.  The demonstration that this formalism correctly captures the singular terms at NLO for the Higgs+1-jet cross section was performed in Refs.~\cite{Liu:2012sz,Liu:2013hba}.  We have confirmed that this is also true for $Z$+1-jet. 
 
We must now correct for the fact that the ATLAS measurement does not impose a global veto on a second jet with $\pTjet > \pTcut$, and instead accepts such dijet events when the second jet falls within a cone around either lepton.  Such events occur first in fixed-order perturbation theory in processes with two final-state partons emitted along with the $Z$-boson.  They can therefore be incorporated in our framework by using the fixed-order result with the ATLAS isolation criterion instead when matching in Eq.~(\ref{rgimproved}):
\bea\label{eq:sigtotal}
\sigma^{\text{NLL}^{\prime}+\text{NLO}}_{\text{total}}= \sigma^{{\rm NLL}'}_{\text{Z+1j}} + \,
\sigma^{\rm NLO}_{\text{total}}-\sigma^{{\rm NLL',exp}}_{\text{Z+1j}}.
\eea
This expression incorporates both the full NLO result for the ATLAS isolation criterion and the resummation of the global jet-veto logarithms, and is our final prediction.  
 
Since it is relevant for our understanding of the numerical results in a later section, we briefly discuss the partonic channels that contribute to the $Z$+1-jet cross section.  At leading order the contributing partonic channels are $q\bar{q} \to Zg$ and $qg \to Zq$, where for the second process the quark can also be an anti-quark.  At NLO, the $gg \to Zq\bar{q}$ and $qq \to Zqq$ also enter as real-radiation corrections.  The SCET framework incorporates the $gg$ and $qq$ initial states in two places: through collinear splittings in the beam-function matching coefficients ${\cal I}_{a,i_aj_a}$, and through the matching to fixed order.  However, in the collinear limit described by the SCET framework these channels necessarily consist of a high-$p_T$ $Z$-boson recoiling against a single jet.  This will turn out to be a bad approximation at high-$\pTjet$ for the ATLAS isolation criterion.  The matching corrections from the $qg$ and $qq$ channels will become extremely large, and will dominate the prediction at high $\pTjet$ for the ATLAS isolation criterion.

One final issue which we discuss briefly is the effect of electroweak Sudakov logarithms.  These arise from electroweak corrections involving $W$ and $Z$ bosons, and lead to another potentially large shift that grows with increasing $\pTjet$ and $p_{T}^{Z}$.  They have been studied previously~\cite{Denner:2011vu,Becher:2015yea}, and were found to lead to corrections which can reach $-20\%$ for transverse momenta around 1 TeV.  As our focus here is on the interplay between different large sources of QCD corrections we will not discuss them further, but they should be included in complete predictions for this process.

\section{Numerical results}
\label{sec:numerics}

We present here and discuss in detail numerical results for 13 TeV LHC collisions.  We use CTEQ parton distribution functions (PDFs)~\cite{Lai:2010vv} at the appropriate order in perturbation theory: LO PDFs for the LO fixed-order cross section, and NLO PDFs for the NLO fixed-order cross section and for our resummed cross sections.  As we will study several different theoretical predictions, we begin with a brief description of the terminology that we will use for our results.
\begin{itemize}

\item {\it Fixed order:} this is the standard result of fixed-order perturbation theory at either LO or NLO, obtained using MCFM.  Unless noted otherwise, the scale choice $\mu_R=\mu_F=H_T/2$ is taken, where $H_T$ is the scalar sum of the transverse momenta of all jets and leptons in the final state.

\item {\it Resummed:} this is the cross section implementing the NLL$^{\prime}$ resummation of Eq.~(\ref{factgen}), but without matching to fixed order.  It therefore includes the resummation of the global jet-veto logarithms.

\item{\it Matched:} this is the full NLL$^{\prime}$+NLO cross section of Eq.~(\ref{eq:sigtotal}).  It is our ``best" prediction that contains the most information about the perturbative expansion.

\end{itemize}
These cross sections will each reveal different important aspects of the perturbative cross section.

\subsection{Results for the ATLAS isolation criterion}
\begin{figure}[!h]
\begin{center}
\includegraphics[width=0.72\textwidth,angle=0]{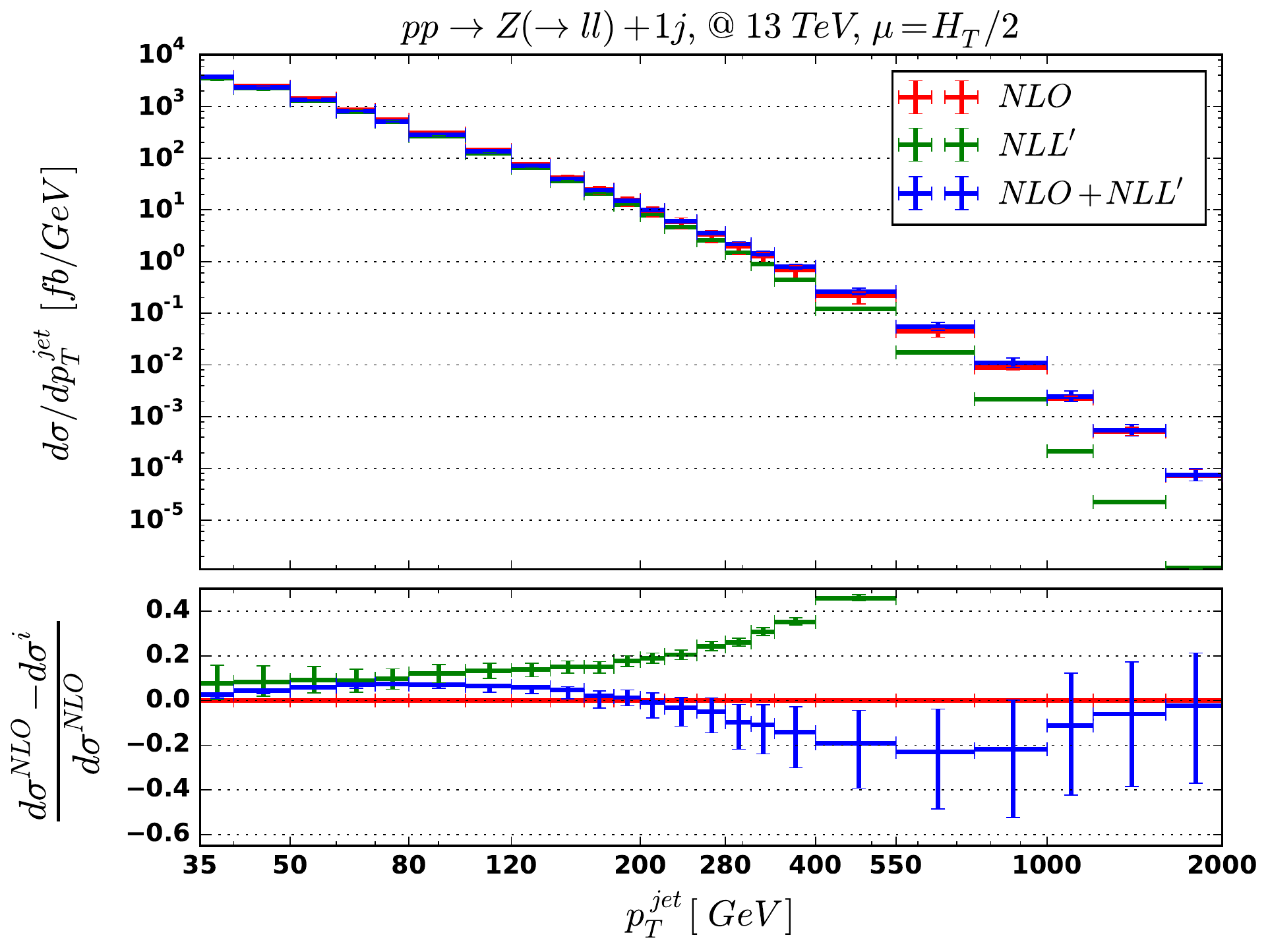}
\end{center}
\vspace{-0.1cm}
\caption{Comparison of the resummed, matched and fixed order spectra (upper panel) and the relative deviations of the resummed and matched predictions with respect to the fixed-order result (lower panel) at 13 TeV for the ATLAS isolation criterion.} \label{fig:13TeVresmatch}
\end{figure}

We begin our discussion with a comparison of the NLO fixed-order result, the resumed result of  Eq.~(\ref{factgen}) and the matched result of Eq.~(\ref{rgimproved}), as shown in Fig.~\ref{fig:13TeVresmatch}, using exactly the ATLAS setup.  We have obtained our fixed-order result using MCFM.  We note that the horizontal error bars indicate the $\pTjet$ bin width, while the vertical ones denote the scale-variations of the theoretical predictions.

The comparison of the fixed-order and matched predictions reveals an interesting structure.  The two results differ by roughly 5\% in the intermediate range  $\pTjet \approx 50-120$ GeV, but agree almost identically for $\pTjet \approx 160-220$ GeV.   Within the substantial scale-variation uncertainties, the matched and fixed-order results agree as well for $\pTjet > 1$ TeV.   
This is not the expected behavior if the jet-veto logarithms dominate the theoretical prediction; they should increase as the ratio $\pTjet/\pTcut$ is increased, and their resummation should decrease their effect on the cross section. 
The resummed prediction is reduced by nearly 50\% with respect to the fixed-order NLO prediction at $\pTjet$ values around 550 GeV, and becomes more than an order of magnitude smaller at $\pTjet \approx 2$ TeV. The matched result on the other hand agrees to better than 2\% with the fixed order result at 
$\pTjet = 2$ TeV. Such a large correction when going from the resummed to the matched prediction indicates that another effect besides the jet-veto logarithms dominates at high $\pTjet$.   

\begin{figure}[!h]
\begin{center}
\includegraphics[width=0.72\textwidth,angle=0]{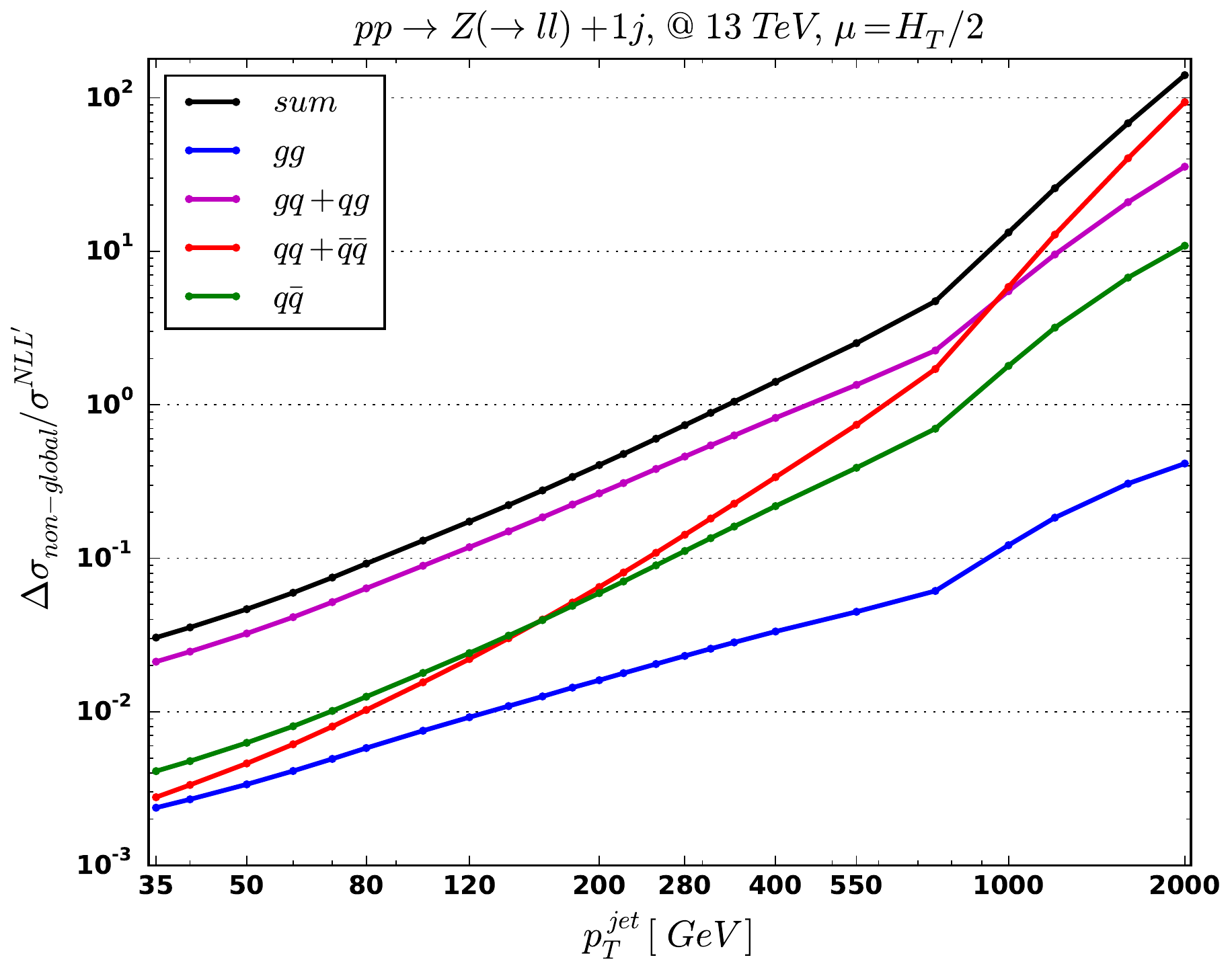}
\end{center}
\vspace{-0.1cm}
\caption{Ratio of the correction factor $\Delta \sigma_{\text{non-global}}$ separated into initial-state partonic channels over the resummed cross section as a function of $\pTjet$.} \label{fig:13TeVpartonic}
\end{figure}

\begin{figure}[!h]
\begin{center}
\includegraphics[width=0.72\textwidth,angle=0]{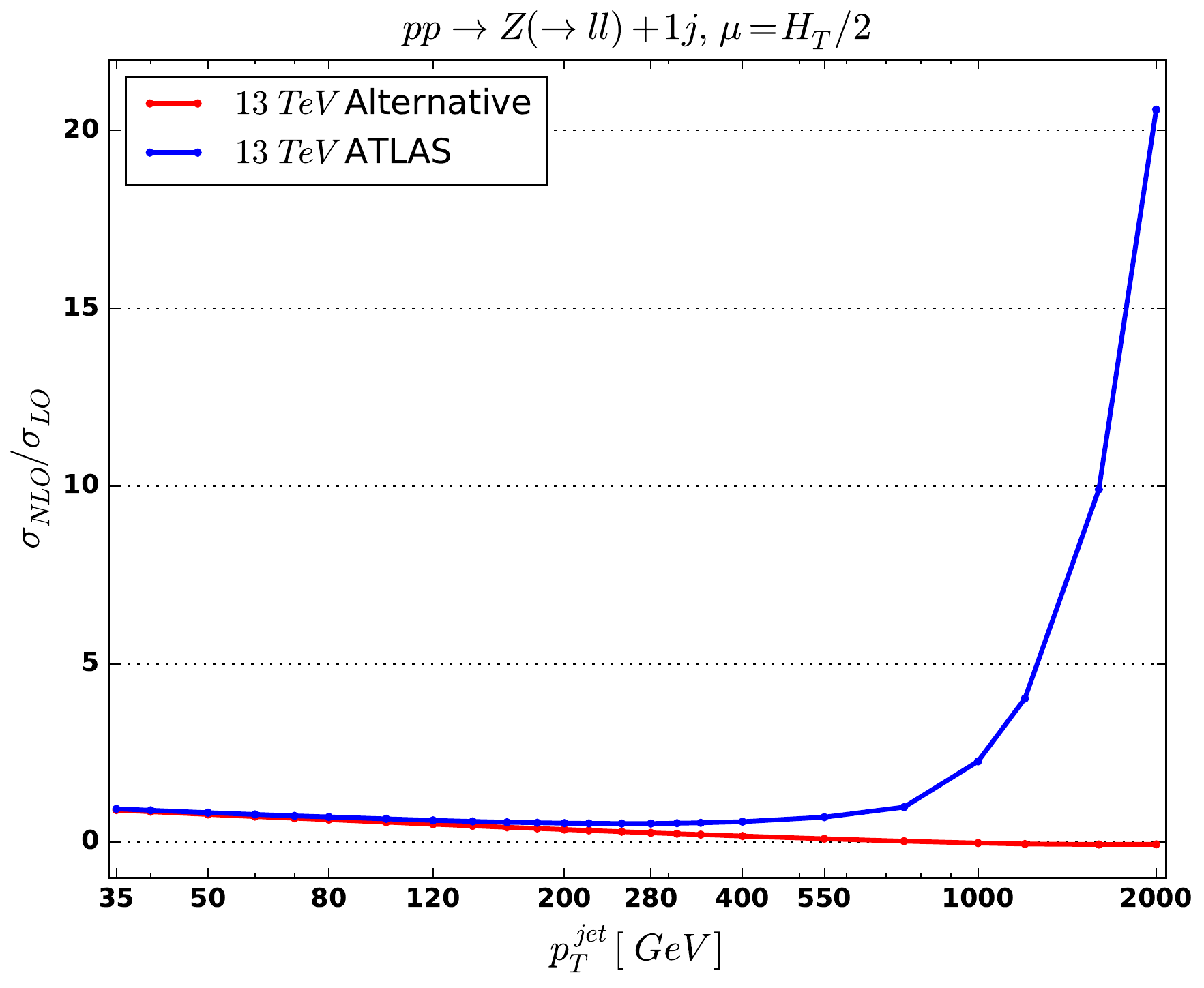}
\end{center}
\vspace{-0.1cm}
\caption{Ratio of the NLO fixed-order cross section over the LO result for 13 TeV for the ATLAS isolation criterion as well as the alternate isolation criterion. } \label{fig:Kfac13}
\end{figure}

The explanation for this effect becomes clear when considering the fact that ATLAS does not impose a global jet veto, but instead allows two-jet events where one jet is collinear to a lepton.  We define this correction factor as $\Delta \sigma_{\text{non-global}} = \sigma^{\rm NLO}_{\text{total}}-\sigma^{\rm NLO}_{\text{Z+1j}}$.  This is exactly the difference between the Z+1-jet cross section and the total cross section defined in Eqs.~(\ref{rgimproved}) and~(\ref{eq:sigtotal}), respectively.  We separate this correction factor into the various initial-state partonic channels, and plot their ratios with respect to the resummed cross section in Fig.~\ref{fig:13TeVpartonic}.  The contribution from two-jet events becomes of the same order as the resummed cross section for $\pTjet \approx 900$ GeV, and overwhelms the resummed result for higher $\pTjet$ values.  The reason for this large effect is that these two-jet events are effectively a dijet process with the emission of a collinear $Z$-boson.
It is an example of a giant $K$-factor, as discussed in Ref.~\cite{Rubin:2010xp}.  We note that the largest effects are in the $qg$ and $qq$ partonic channels, due to their large luminosities at high Bjorken-$x$.  Further evidence for the dominance of this new kinematic configuration is provided by the ratio of the NLO fixed-order result over the LO cross section, shown in Fig.~\ref{fig:Kfac13} for 13 TeV collisions.  The ratio for the ATLAS isolation criterion is below one for $\pTjet$ up to approximately 700 GeV, but grows to over 30 for the highest $\pTjet$ values shown.  We have shown in this plot the ratio of the NLO over the LO cross section for the alternate isolation criterion discussed in Section~\ref{sec:atlas}, for which these dijet events are removed.  It does not show a similar dramatic increase at high $\pTjet$, further confirming the origin of this large correction.

To further clarify in detail the cancellation between the ``giant K-factor" effects related to the dijet events and the jet-veto logarithms in the  $\pTjet$ range of 1000-1600 GeV, 
we perform the following analysis.  
We rewrite the resummation improved prediction at NLL'+NLO using ATLAS isolation cuts, denoted in the following as 
$ \sigma_{ATLAS}^{matched}$, in the following form:  
\begin{equation}
   \sigma_{ATLAS}^{matched} =  \sigma^{NLO}_{alternate}+ \{ \sigma^{matched}_{alternate} - \sigma^{NLO}_{alternate}\} +  
   \{ \sigma^{NLO}_{ ATLAS} - \sigma^{NLO}_{alternate}\}\,.
\end{equation}

The label alternate refers to our alternate isolation criterion in which jets with $\Delta R_{lj} \leq 0.5$ are vetoed, which means that dijet events are removed and only events with exclusive $Z+1j$ final state are included. This equation is an identity. The way to view it is the following.
\begin{itemize}
\item We begin with the $\sigma^{NLO}_{alternate}$ prediction.
\item We add on the first bracket, which accounts for the resummation of jet-veto logarithms using our formalism; it is the difference of the matched result and the NLO result, both using the alternate isolation.
\item Finally, we add on the second bracket, which is the difference between the NLO result in the ATLAS isolation and our alternate isolation.  This bracket adds on the dijet events responsible for the giant K-factor\,.
\end{itemize}
Below we show the numerical results for the first and second bracket  in the kinematic region $1000$ GeV $< p_{T}^{jet}<1600$ GeV and for the central scale choice $\mu=H_T/2$:
\begin{eqnarray}
&&\sigma^{NLO}_{alternate} =  -0.0071 \, \text{fb}\\
&&\sigma^{NLO}_{ ATLAS}   \;\;=  0.6485 \, \text{fb} \\
&&\{ \sigma^{matched}_{alternate} - \sigma^{NLO}_{alternate}\} =  0.06159 \, \text{fb}\\
 &&\{ \sigma^{NLO}_{ ATLAS} - \sigma^{NLO}_{alternate}\}  \;=  0.6555 \, \text{fb}
\end{eqnarray}
A few aspects of the results are visible from these numbers.  The first is that the effect of jet-veto logarithms on the exclusive $Z$+1-jet cross section is so large that the fixed-order prediction $\sigma^{NLO}_{alternate}$ is negative. The resummation of these logarithms, quantified by the difference 
$\{ \sigma^{matched}_{alternate} - \sigma^{NLO}_{alternate}\}$, is needed to make this cross section positive.  However, this effect 
is masked with the ATLAS isolation criterion by the large contribution from dijet events, shown above in the difference 
$\{ \sigma^{NLO}_{ ATLAS} - \sigma^{NLO}_{alternate}\}$.  If the alternate isolation can be experimentally investigated, it would offer 
a direct test of the jet-veto resummation formalism.

\subsection{Results for the alternate isolation}

One important application of the exclusive $Z$+1-jet measurement is to test the jet-veto resummation framework that promises to have a large impact on Run II Higgs analyses.  From that perspective the above result is disappointing, since the effect of jet-veto resummation is overwhelmed by the giant $K$-factor.  Although there is a $\pTjet$ region in 13 TeV collisions in which there is a roughly 25\% difference between the fixed-order result and the full matched result, this arises from a cancellation between the two large effects in the perturbative expansion, and may not be stable with respect to unknown higher-order corrections.  However, there is a way around this problem.  The giant $K$-factor comes from the phase-space region where the $Z$-boson is emitted collinear to an energetic jet.  This leads to a large positive correction.  This phase-space region can be removed by adopting the alternate isolation criterion discussed in Section~\ref{sec:atlas}.  The giant $K$-factor no longer appears if the collinear emission of the jet along the $Z$-boson direction is vetoed.  To verify this we plot in Fig.~\ref{fig:Kfac13} the NLO over LO $K$-factors for 13 TeV collisions using the alternate isolation criterion.  The large $K$-factor at high $\pTjet$ is no longer present, as expected. The cross section in 13 TeV collisions even becomes negative at high $\pTjet$ due to the large jet-veto logarithms present in the fixed-order result.  A more detailed experimental investigation is needed to determine whether the adoption of this isolation criterion is possible.  However, since it amounts only to modifying the Monte Carlo correction applied to the data as discussed in Section~\ref{sec:atlas}, we are confident that some variant of this proposal should be possible.

\begin{figure}[!h]
\begin{center}
\includegraphics[width=0.72\textwidth,angle=0]{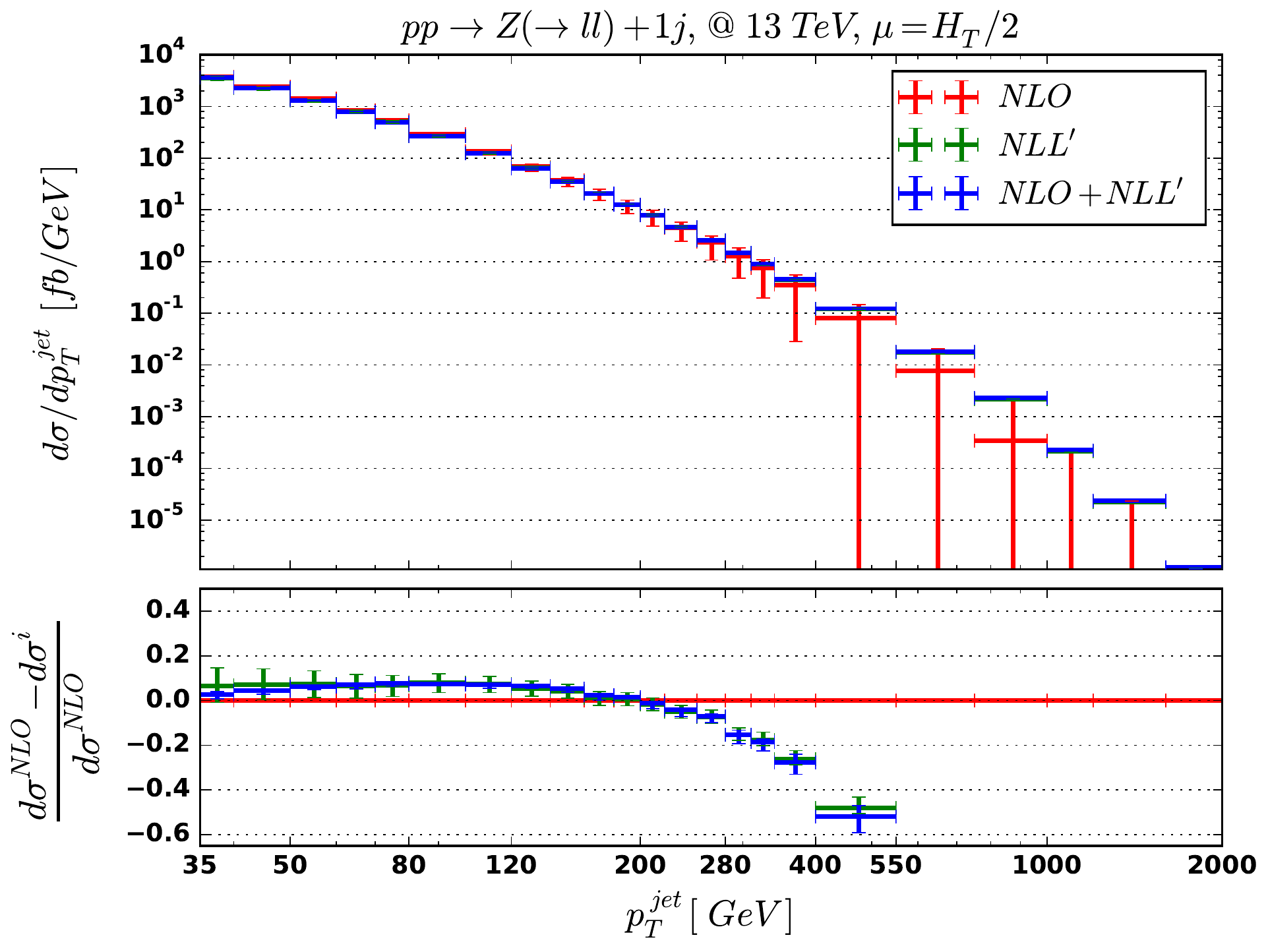}
\end{center}
\vspace{-0.1cm}
\caption{Comparison of the fixed-order NLO, resummed and matched results in 13 TeV for the alternate isolation criterion.  The lower inset shows the relative deviations of the theoretical predictions with respect to fixed-order.} \label{fig:alteriso}
\end{figure}

To study what may be learned from investigating the alternate isolation criterion with 13 TeV data, we show in Fig.~\ref{fig:alteriso} the comparisons of fixed-order with the resummed and matched results.  We will focus our explanation on the high-$\pTjet$ region, where we expect the largest discrepancies between fixed-order and the resummation formalism to occur.  We first note that the resummed prediction and the matched result are nearly the same for all $\pTjet$ values.  There is no longer a large correction to the resummed cross section as there was with the ATLAS isolation criterion.  The deviation between fixed-order and the matched result reaches 50\% at $\pTjet \approx 500$ GeV.  The discrepancy becomes even larger for higher $\pTjet$, when the fixed-order result becomes negative as seen in Fig.~\ref{fig:Kfac13}.  From the perspective of testing the jet-veto resummation formalism, these are exactly the desired results: large discrepancies with respect to fixed-order predictions in kinematically accessible phase-space regions.  Measurement of the high-$\pTjet$ cross section with the alternate isolation criterion suggested here will therefore provide a strong test of the resummation formalism.

\begin{figure}[!h]
\begin{center}
\includegraphics[width=0.72\textwidth,angle=0]{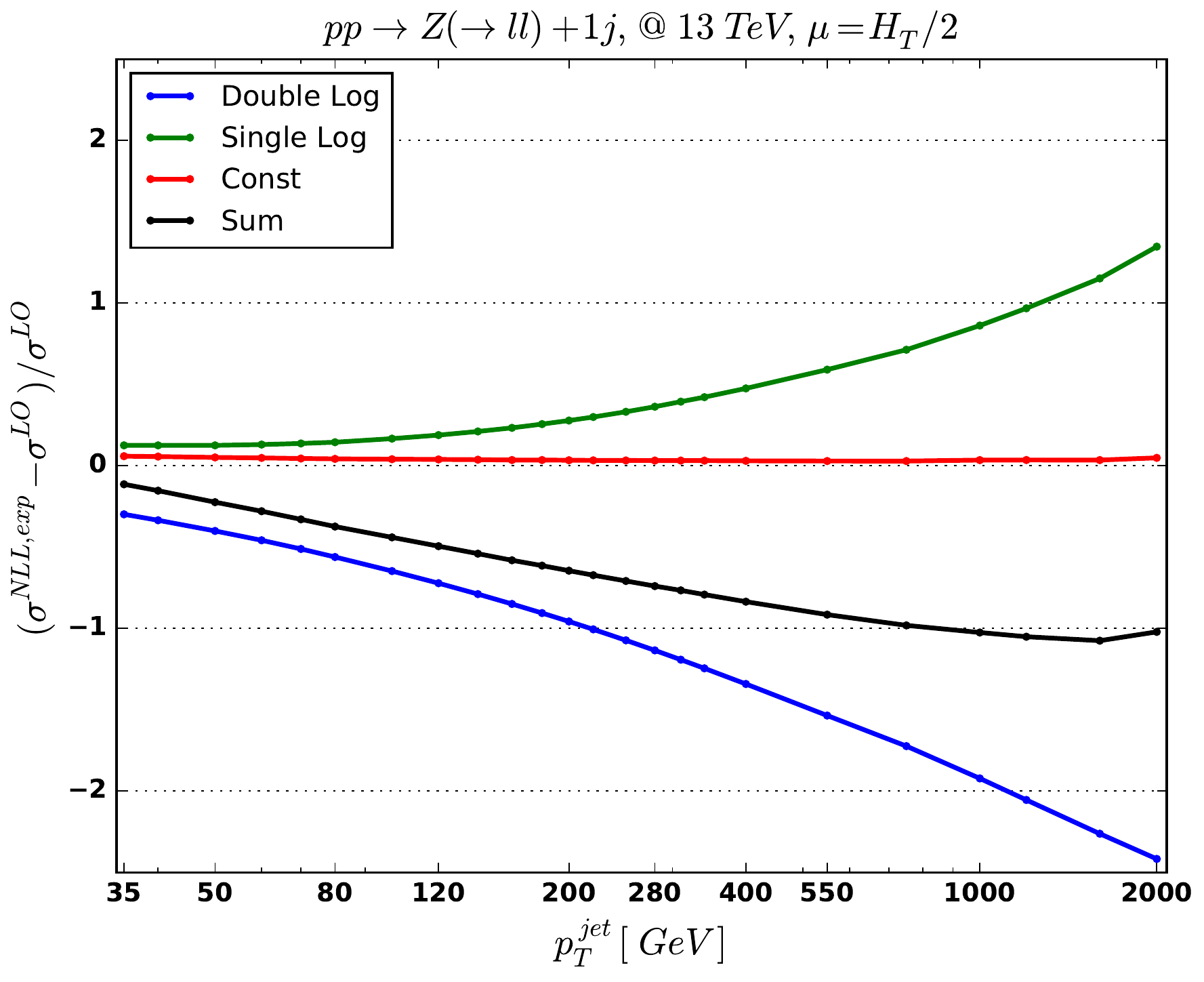}
\end{center}
\vspace{-0.1cm}
\caption{Comparison of the leading double logarithms, single logarithms, and constant contributions to the exclusive cross section at relative order $\alpha_s$ with respect to the leading order.  All the contributions are normalized to the LO cross section.  
} \label{fig:LLNLL}
\end{figure}

Finally, we point out one other interesting aspect of the jet-veto logarithms that shows the importance of a precise treatment of these effects in theoretical predictions.  We expand the resummation-improved result of Eq.~(\ref{factgen}) to NLO in $\alpha_s$, and study separately the effects of the leading double logarithms, the single logarithms, and the constant terms that appear in the cross section.  All power-suppressed terms in $\pTcut$ are dropped in this expansion.  Fig.~\ref{fig:LLNLL} shows that there is a cancellation between the leading-log and single-log terms, which reduces the effect of the logarithmic corrections at intermediate and high $\pTjet$.  A leading-logarithmic estimate of the region of $\pTjet$ in which jet-veto logarithms dominate the theoretical prediction would therefore underestimate the $\pTjet$ value for which this occurs.  The cancellation effectively postpones the breakdown of fixed-order perturbation theory for this process.

\subsection{Other observables}

Finally, we discuss the possibility of other observables that exhibit sensitivity to jet-veto resummation.  One possibility is the transverse momentum distribution of the $Z$-boson.  If this is measured in the exclusive one-jet bin at high-$\pTZ$, it will exhibit similar large logarithmic corrections to the $\pTjet$ distribution.  Furthermore, the requirement of a highly-energetic $Z$-boson reduces the possibility of a soft $Z$-boson arising from an underlying dijet configuration, reducing the sensitivity to new scattering channels that appear at higher orders.  As was apparent from the kinematical considerations of Section~\ref{sec:kinematics}, there is a significant contribution from dijet events with $\pTjet  > \pTZ$.  Demanding a high transverse momentum $Z$-boson removes these events, increasing the impact of the jet-veto logarithms.

\begin{figure}[!h]
\begin{center}
\includegraphics[width=0.72\textwidth,angle=0]{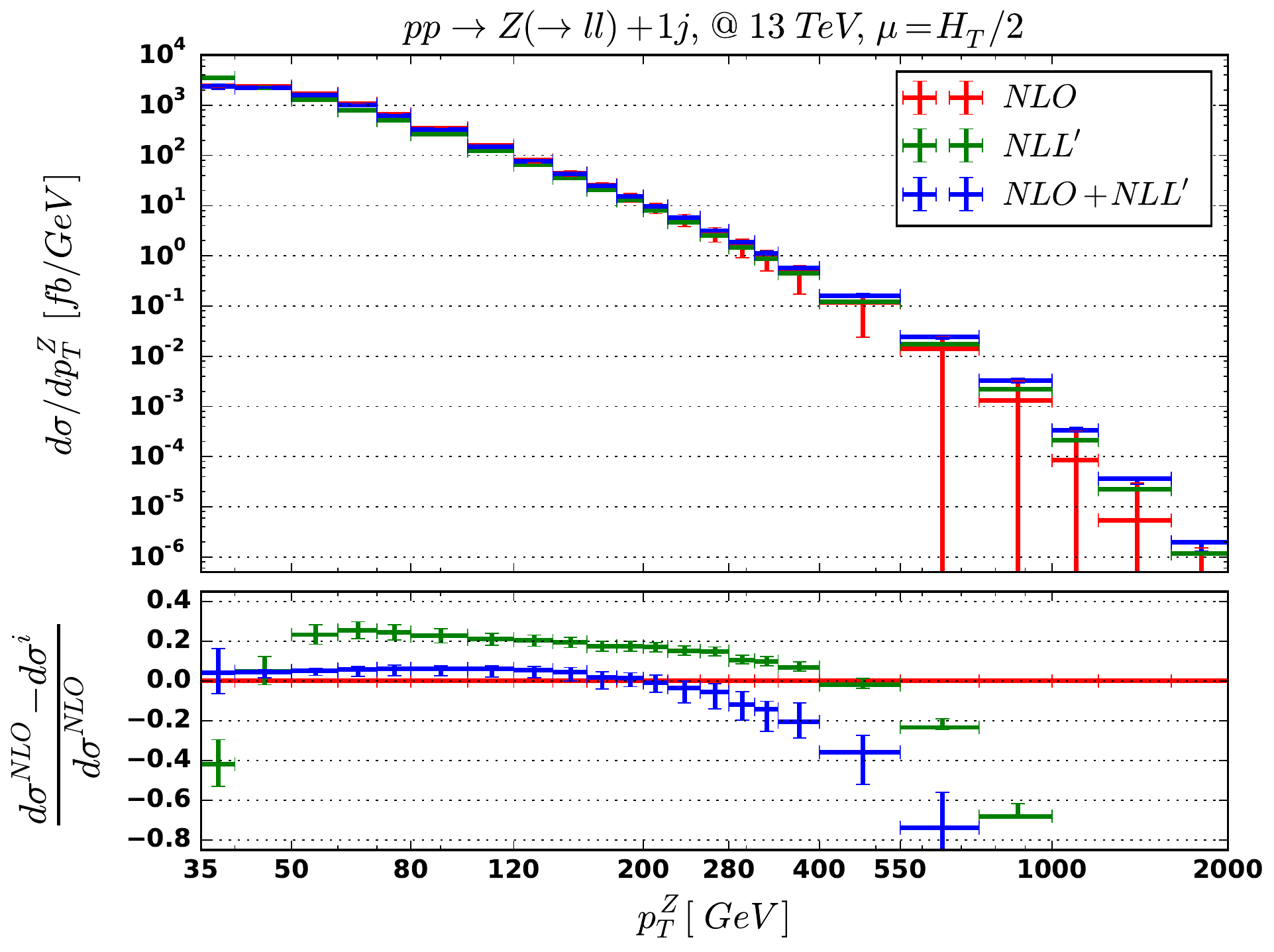}
\end{center}
\vspace{-0.1cm}
\caption{Comparison of the resummed, matched and fixed order spectra for the $Z$-boson transverse momentum (upper panel) and the relative deviations of the resummed and matched predictions with respect to the fixed-order result (lower panel) at 13 TeV for the ATLAS isolation criterion.} \label{fig:13TeVPTZatlas}
\end{figure}
\begin{figure}[!h]
\begin{center}
\includegraphics[width=0.72\textwidth,angle=0]{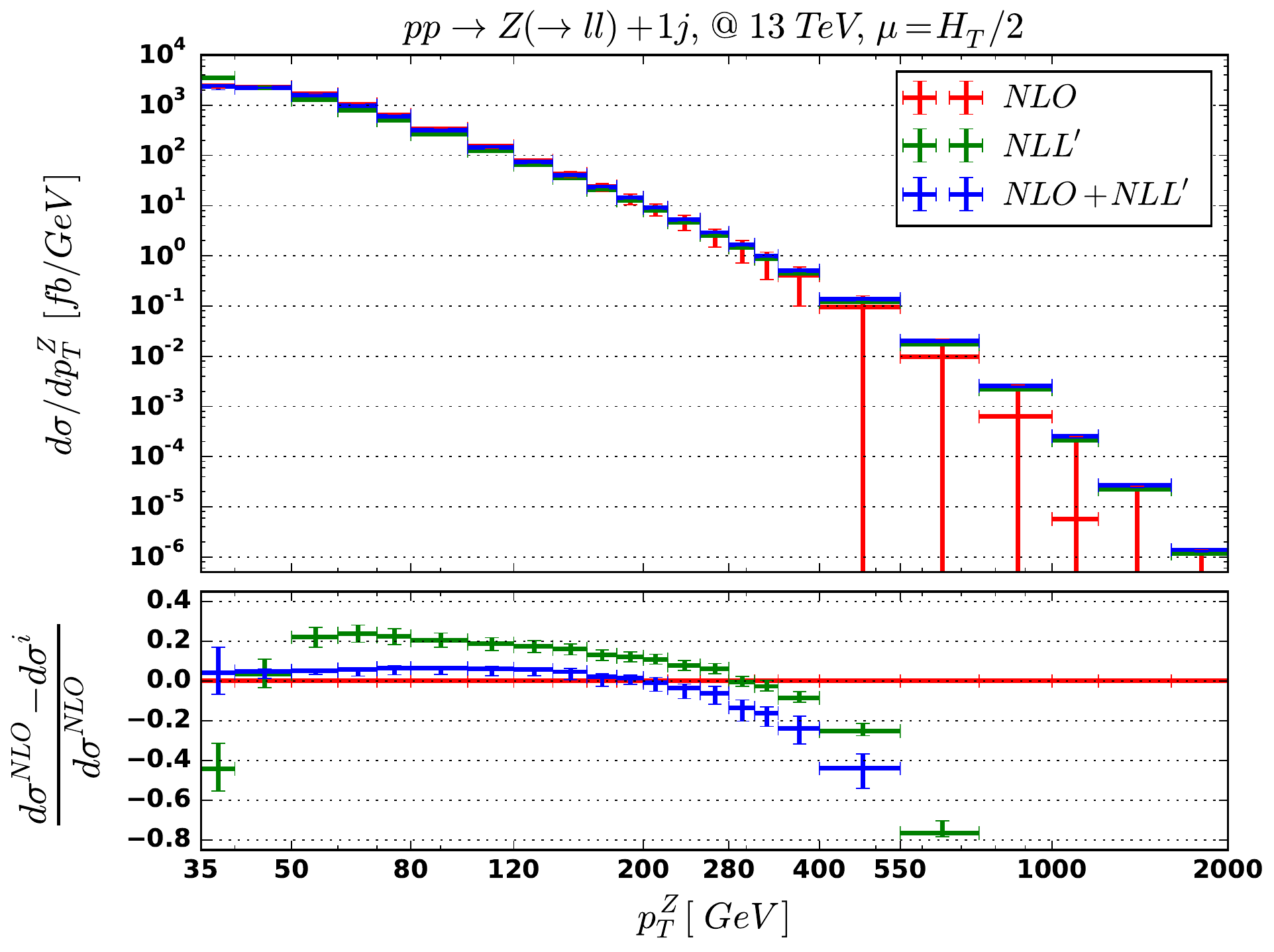}
\end{center}
\vspace{-0.1cm}
\caption{Comparison of the resummed, matched and fixed order spectra for the $Z$-boson transverse momentum (upper panel) and the relative deviations of the resummed and matched predictions with respect to the fixed-order result (lower panel) at 13 TeV for the alternate isolation criterion. } \label{fig:13TeVPTZiso}
\end{figure}

We show in Figs.~\ref{fig:13TeVPTZatlas} and~\ref{fig:13TeVPTZiso} the $\pTZ$ spectra for both the ATLAS and alternate isolation criteria, respectively.  We compare the fixed-order NLO results with those from the NLL$^{\prime}$ resummation and the full NLL$^{\prime}$+NLO matched results.  For both isolation criteria there is a significant difference between the NLO result and the NLL$^{\prime}$ resummed cross section for $\pTZ > 500 $ GeV, indicating that jet-veto logarithms have an important effect on this observable.  The full matched result is closer to the NLL$^{\prime}$ for the alternate isolation criterion than for the ATLAS choice. However, in both cases the effect of jet veto resummation will be observable in the data at high-$\pTZ$.

\section{Conclusions}
\label{sec:conc}

In this paper we have studied in detail the exclusive $Z$+1-jet cross section at a 13~TeV LHC.   This measurement in principle should test the theoretical framework for jet-veto resummation that promises to greatly reduce the uncertainties plaguing the interpretation of Higgs-boson analyses in the $WW$ final state.  We have adopted the experimental cuts used at 7 TeV by the ATLAS collaboration.  We have identified an important aspect of the ATLAS analysis that makes it difficult to test the resummation formalism in this process: the ATLAS cuts do not apply a global veto on a second jet, but instead allow such an additional jet to be collinear to one of the final-state leptons.  The result in this region of phase space contains a giant $K$-factor arising from the emission of a collinear $Z$-boson from an underlying dijet process.  This configuration dominates the cross section at high $\pTjet$.  We have provided numerical predictions that account for both the resummation of jet-veto logarithms and the giant $K$-factor, and have studied the interplay between these competing effects.
 
The isolation criterion implemented by ATLAS mixes the effect of the jet-veto resummation framework with the giant $K$-factor.  It is desirable to find a way to isolate the jet-veto logarithms in the perturbative expansion, in order to test the resummation formalism.  We have therefore suggested an alternate criterion that imposes a global veto on a second jet with $\pTjet > \pTcut$, thereby removing the giant $K$-factor.  We have provided numerical results using this global veto, and have demonstrated that the jet-veto resummation now dominates the theoretical predictions  We have further studied the possibility of testing the resummation formalism using the $\pTZ$ distribution.  Focusing on the high-$\pTZ$ region naturally removes dijet events with a soft $Z$-boson, thereby reducing the giant $K$-factor effect even for the ATLAS isolation criterion.

We encourage the experimental collaborations to measure exclusive $Z$+jet cross sections using both isolation criteria discussed in this paper.  We look forward to testing the jet-veto resummation framework with this data.
 
\section{Acknowledgments}
\label{sec:acks}

We thank Joey Huston for many useful communications. We are grateful to Daniel Maitre for providing cross checks of our 7 TeV fixed-order predictions. We are indebted to Sergei Chekanov for providing local computing resources. The work of R.~B. was supported by the U.S. Department of Energy, Division of High Energy Physics, under contract DE-AC02-06CH11357.  The work of C.~F. was supported by the U.S. Department of Energy, Division of High Energy Physics, under the grant DE-SC0010143.  The work of X.~L. was supported by the U.S. Department of Energy.

\appendix
\section*{Appendix}
In this Appendix we list some of the ingredients used in the NLL$^{\prime}$ resummation.
\section{Jet Function}
For the partonic channel $qg \to q Z $, we need the quark jet function up to one-loop order:
\bea
J_{q} = 1 + \frac{\alpha_s(\mu)}{4\pi}  \left[
\Gamma_0 T_q^2 L^2 + \gamma_0^{J_q} L + \left( 13 - \frac{3\pi^2}{2} \right) C_F
\right] \,,
\eea
where $L =  \log \frac{\mu}{p_T^{jet} R}$. \\
For $q{\bar q} \to gZ$, we require the one-loop gluon jet function:
\bea
J_{g} = 1 + \frac{\alpha_s(\mu)}{4\pi}  \left[
\Gamma_0 T_g^2 L^2 + \gamma_0^{J_g} L \,
+ \left(\frac{134}{9} - \frac{3\pi^2}{2} \right) C_A - \frac{23}{9}n_f 
\right] \,.
\eea
We note that
\bea
 &&\Gamma_0 = 4\,, \quad \quad
 \gamma_0^{J_q} = 6C_F \,, \quad \quad  \gamma_0^{J_g} = 2\beta_0 \,.
\eea
%
\section{Beam Function}
The beam function can be written as a convolution,
\bea
B_i(x) = f_i(x) +  \sum_j \int_x^1 \frac{\mathrm{d} z}{z} \, 
{\cal I}^{(1)}_{ij}\left( z \right) f_j\left( \frac{x}{z}\right)  + \cdots \,,
\eea
where the NLO matching coefficients ${\cal I}_{ij}$ are found to be
\bea
{\cal I}^{(1)}_{gg}(z) &=& \,
\frac{\alpha_s(\mu)C_A}{2\pi}\left(\,
4\log\frac{\mu}{\pTcut}\log\frac{\nu}{\bnp} \delta(1-z)
-2\tilde{p}_{gg}(z) \log \frac{\mu}{\pTcut} 
 \right)\,, \nn \\
{\cal I}^{(1)}_{qq}(z) &=& \,
\frac{\alpha_s(\mu)C_F}{2\pi}\left(\,
4\log\frac{\mu}{\pTcut}\log\frac{\nu}{\bnp}\delta(1-z)\,
-2\tilde{p}_{qq}(z) \log \frac{\mu}{\pTcut}\,
+(1-z)
\right)\,, \nn \\
{\cal I}^{(1)}_{gq}(z) &=& \frac{\alpha_s(\mu)C_F}{2\pi}\left(\,
-2p_{gq}(z)\log\frac{\mu}{\pTcut} + z 
\right)\,, \nn \\
{\cal I}^{(1)}_{qg}(z) &=& \frac{\alpha_s(\mu)T_F}{2\pi}\left(\,
-2p_{qg}(z)\log\frac{\mu}{\pTcut} + 2z(1-z) 
\right) \,,
\eea
with 
\bea
&&\tilde{p}_{gg}(z) = \frac{2z}{(1-z)_+}+2z(1-z)+2\frac{1-z}{z}\,, \nn\\
&&\tilde{p}_{qq}(z) = \frac{1+z^2}{(1-z)_+}  \,, \nn \\
&&p_{gq}(z) = \frac{1+(1-z)^2}{z} \,, \nn \\
&&p_{qg}(z) = 1 -2z +2z^2\,.
\eea
The $+$-prescription is implemented via
\bea
\int_x^1 \frac{\mathrm{d} z}{z} \, \left(\frac{1}{1-z} \right)_+ \, f\left(\frac{x}{z}\right) F(z) 
&=& \int_x^1  \mathrm{d} z \, \frac{1}{1-z} \left( f\left(\frac{x}{z}\right) F(z)  \frac{1}{z}  - f(x) F(1)   \right) \nn \\
&&+ f(x) F(1) \log(1-x) \,.
\eea
%

\section{Soft Function}
The one-loop soft function is found to be
\bea
S = 1 + \frac{\alpha_s}{4\pi}\left(\sum_{a\in B}T_a^2 \,
\left[L^2 + 4\log\frac{\pTcut}{\nu}L \right] 
+ 2T_J^2\log R^2 L\,
+ 4(T_a\cdot T_J - T_b\cdot T_J)\, y_J L
+c_S\right)
\eea
with $L = \log(\mu/\pTcut)^2$, and
\bea
c_S = -\left( T_a^2+T_b^2\right) \frac{\pi^2}{6} \,
+ T_J^2 \,\left(9.22045+ f(R)\right)\,,
\eea
where
\bea
f(R) = -4\log(2) \log R^2 \,
+8\int_{-\infty}^\infty \mathrm{d} \Delta y \,
\int_0^\pi\frac{\mathrm{d}\Delta \phi}{\pi} \, 
\frac{\log(s_{\Delta \phi})}{\Delta {R_{kJ}}^2}\,
\Theta_{\Delta {R_{kJ}},R}  \,.
\eea
We have $f(0.4) =-12.5778$ and $f(0.5) = -11.1423$. 
For moderate $R\sim {\cal O}\left(10^{-1}\right)$, we can approximate
\bea
f(R)
 = -9.22352 -0.00219773 \log R^2 - \log^2 R^2
\,.
\eea
%

%
%
\section{Evolution}

The evolutions of the jet and the beam functions are given by
\bea
U_{J_i}(\mu_J,\mu) &=& \exp\left[-2T_i^2 S(\mu_J,\mu)-A_{J_i}(\mu_J,\mu) \right]
\left( \frac{\mu_J}{p_T^{J}R}\right)^{-2T_J^2 A_\Gamma(\mu_J,\mu)} \,, \nn \\
U_{B,a}(\mu_B,\nu_B,\mu,\nu) &=& 
\exp\left[-T_a^2 A_\Gamma(\pTcut,\mu)\log\frac{\nu^2}{\nu_B^2}\right]
\exp\left[-T_a^2 A_\Gamma(\mu_B,\mu)\log\frac{\nu^2_B}{\w_a^2}-A_{B_a}(\mu_B,\mu) \right]\,. \nn \\
\eea
The solution to the RG equation for the hard function is
\bea
U_H (\mu_H,\mu)&=&\exp\left[2\sum_i T_i^2 S(\mu_H,\mu)-2A_H(\mu_H,\mu)\,
+2A_\Gamma(\mu_H,\mu) \sum_{i\neq j}\frac{T_i\cdot T_j}{2}\log\Delta R^2_{ij}\right]\nn\\
&& \times
 \prod_i \left(\frac{\mu_H}{\w_i} \right)^{2T_i^2A_\Gamma(\mu_H,\mu)}\,, 
\eea
where we have set $\Delta R^2_{Ja} = e^{-\eta_J}$, $\Delta R^2_{Jb} = e^{\eta_J}$ and $\Delta R^2_{ab} = 1$. We also set $\w_i = p_T^{J}$ if $i = J$; otherwise we have $\w_a = x_a\sqrt{s}$.  The soft-function evolution factor is
\bea
U_S(\mu_S,\nu_S,\mu,\nu)& = &\exp\left[-2\sum_{i\in B}T_i^2 S(\mu_s,\mu) -A_s(\mu_s,\mu)\, 
- 2A_\Gamma(\mu_s,\mu)\sum_{i \ne j}\frac{T_i\cdot T_j}{2}\log \Delta R^2_{ij}\right]\nn\\
&& \times \left(\frac{1}{R}\right)^{2T_J^2A_{\Gamma}(\mu_s,\mu)}
\left(\frac{\nu_s}{\mu_s} \right)^{\sum_{i\in B}2T_i^2A_{\Gamma}(\mu_s,\mu)}
\left(\frac{\nu}{\nu_s} \right)^{\sum_{i\in B}2T_i^2A_{\Gamma}(\pTcut,\mu)}\,.
\eea
For the NLL$^{\prime}$ resummation, we need the following factors:
\bea
A_{\Gamma}(\mu_i,\mu_f)&=& \frac{\Gamma_0}{2\beta_0}\left\{
\log r + \,
\frac{\alpha_s(\mu_i)}{4\pi}\left(\frac{\Gamma_1}{\Gamma_0}-\frac{\beta_1}{\beta_0} \right)\,
(r-1) \,
\right\}\,,
\eea
and 
\bea
S(\mu_i,\mu_f) &=& \frac{\Gamma_0}{4\beta_0^2}\left\{ 
\frac{4\pi}{\alpha_s(\mu_i)}\left(1-\frac{1}{r}-\log r\right)\,
+\left(\frac{\Gamma_1}{\Gamma_0}-\frac{\beta_1}{\beta_0} \right)(1-r+\log r)\,
+\frac{\beta_1}{2\beta_0}\log^2 r 
\right\}\,, \nn \\
\eea
where $r=\alpha_s(\mu_f)/\alpha_s(\mu_i)$. 
$A_{J/B}$, $A_H$ and $A_S$ are needed at leading order, and can be obtained
by substituting the $\Gamma_0$ in $A_\Gamma$ with the corresponding
$\gamma_0^i$ and expanding in $\alpha_s$.  
We note that the non-cusp anomalous dimensions of the beam and jet functions are the same:
\bea
\gamma_0^{B_{a,i}} &=& \gamma_0^{B_{b,i}} = \gamma_0^{J_i}\,. 
\eea
We have the following expressions for the necessary anomalous dimensions, as well as the relevant coefficients of the QCD beta functions needed:
\bea
&&\beta_0 = \frac{11}{3} C_A - \frac{4}{3} T_F n_f  \,,\nn \\
&&\beta_1 = \frac{34}{3} C_A^2 - \frac{20}{3} C_A T_F n_f - 4 C_F T_F n_f \, .
\eea
For $\gamma^H = \sum_i \gamma^{H_i}$, we have
\bea
	\gamma_0^{H_q} = -3 C_F\,, \quad \quad \gamma_0^{H_g} = -\beta_0.
\eea
At one loop, $\gamma_0^S = 0$.
The cusp anomalous dimension is given by 
\bea
\Gamma_{cusp}=\frac{\alpha_s}{4\pi}\Gamma_0 + \left(\frac{\alpha_s}{4\pi}\right)^2\Gamma_1 + \ldots
\eea
with
\bea
&&\Gamma_0 = 4 \,,\nn \\
&&\Gamma_1 = 4 \left[ C_A \left( \frac{67}{9} - \frac{\pi^2}{3} \right) - \frac{20}{9} T_F n_f \right]\,.
\eea


\end{document}